\documentclass[a4paper]{article}

\usepackage{icrc2013}
\usepackage[english]{babel}

\title{Typical atmospheric aerosol behavior at the Cherenkov Telescope Array candidate sites in Argentina}

\shorttitle{ICRC 2013, Piacentini et al.}

\authors{
Rub\'en D. Piacentini$^{1,2}$,
Mart\'in Freire $^{1}$,
Mar\'ia I. Micheletti$^{1,3}$,
Graciela M. Salum$^{1,4}$,
Javier Maya$^{5}$,
Alexis Mancilla$^{5}$,
Beatriz Garc\'ia$^{5,6}$,
for the  CTA Consortium.}

\afiliations{
$^1$ \'Area F\'isica de la Atm\'osfera, Radiaci\'on Solar y Astropart\'iculas, Instituto de F\'isica Rosario (CONICET – Universidad Nacional de Rosario), Rosario, Argentina \\
$^2$ Laboratorio de Eficiencia Energ\'etica, Sustentabilidad y Cambio Clim\'atico, Facultad de Ciencias Exactas, Ingenier\'ia y Agrimensura, UNRosario, Rosario, Argentina  \\
$^3$ Facultad de Ciencias Bioqu\'imicas y Farmac\'euticas,  UNRosario, Rosario, Argentina \\
$^4$ Universidad Tecnol\'ogica Nacional, Fac. Regional Concepci\'on del Uruguay, Entre R\'ios, Argentina\\
$^5$ ITEDA (Instituto de Tecnolog\'ias en Detecci\'on y Astropart\'iculas), CNEA,CONICET, UNSAM, Argentina \\
$^6$ Universidad Tecnol\'ogica Nacional, Regional Mendoza, Mendoza, Argentina 
}

\email{ruben.piacentini@gmail.com}

\abstract{Aerosols from natural and antropogenic sources  are one of the atmospheric components that have the largest spacial-temporal variability, depending on the type (land or ocean) surface, human activity and climatic conditions (mainly temperature and wind). Since Cherenkov photons generated by the incidence of a primary ultraenergetic cosmic gamma photon have a spectral intensity distribution concentrated in the UV and visible ranges [Hillas AM. Space Science Reviews, 75, 17-30, 1996], it is important to know the aerosol concentration and its contribution to atmospheric radiative transfer.  We present results of this concentration measured in typical rather calm (not windy) days at San Antonio de los Cobres (SAC) and El Leoncito/CASLEO  proposed Argentinean Andes range sites for the placement of the Cherenkov Telescope Array (CTA). In both places, the aerosol concentration has a peak in the 2.5-5.0$\mu$m range of the mean aerosol diameter and a very low mean total concentration of 0.097$\mu$g/m$^3$ (0.365$\mu$g/m$^3$) for SAC(CASLEO). The data were collected the days 15/Dec/2012 (27-28/Dec/2012) for the first (second) place with a GRIMM aerosol spectrometer, who determines the aerosol concentration  at a given diameter (0.22-32 $\mu$m range) with a laser technique.  We also present AOD values for each CTA proposed place, derived with the improved Deep Blue algorithm, from data measured by SEAWiFS instrument on board of SeaStar/NASA satellite. They have very low mean values, determined for the 1998-2010 period, AOD$_{550nm, SAC}$ = 0.026 $\pm$ 0.011 and AOD$_{550nm, CASLEO}$ = 0.030 $\pm$ 0.014. Also, we introduce the spectral (380 nm) solar irradiance as a test in these sites of the tropospheric UV atmospheric transmittance. 
}

\begin{document}
\maketitle

\section{Introduction}

Cosmic gamma photons and showers of secondary particles
can be detected through its interaction with  atmospheric gases and aerosols.
Among  them, particulate matter  originated from natural and antropogenic
sources are one of the components that have the largest
spacial-temporal variability, depending on the type (land
or ocean) surface, human activity and climatic conditions
(mainly temperature and wind). In particular, Cherenkov photons
generated by the incidence of a primary ultraenergetic cosmic
gamma photon, have a spectral intensity distribution
concentrated in the UV and visible ranges (see Hillas \cite{bib:hillas4}).

In relation to the search for a convenient place for
the previous CTA location (see for example HESS Collaboration \cite{bib:hess5}), Esposito et al. \cite{bib:esposito6},
analyzed the aerosol characteristics at the Khomas highlands in Namibia. In a similar search, in what follows we introduce in item 2 the proposed Argentinean
sites and in items 3.1 and 3.2, the corresponding behavior
of the aerosol concentration (from ground measurements),
aerosol optical depth (from satellite data) and 380 nm solar
spectral irradiance (also from satellite data).
 
The next-generation high-energy
gamma ray observatory, the CTA Consortium (www.cta-observatory.org/) 
will be a wide collaboration from Africa,
Asia, Europe, North and South America. CTA now goes through
its preparatory phase for candidate sites both for the Northern and Southern
hemispheres. These studies, among others, involve atmospheric
quality, according to precise specifications for
the telescopes belonging to CTA \cite{bib:cta1}, \cite{bib:cta2},  \cite{bib:cta3}.
Current systems of Cherenkov telescopes use at most
four telescopes, providing best stereo imaging of particle
cascades over a very limited area, with most cascades
viewed by only two or three telescopes. An array of tens of 
telescopes will allow the detection of gamma-ray
induced cascades over a large area on the ground, increasing
the number of detected gamma rays, while at the same
time providing a much larger number of views of each cascade.
This will result in the improvement of angular resolution
and in a better suppression of cosmic-ray background
events.
The Cherenkov Telescope Array Consortium 
contemplates the design, construction and operation of two
observatories for the detection of gamma-rays produced by
extraterrestrial sources. The observatories will be deployed
in both hemispheres for a full sky coverage. Each observatory
will consist of an array of telescopes, with enough sensitivity
to detect the atmospheric Cherenkov radiation produced
by the gamma-rays,  improving the capacity of the actual facilities. 
In the present design, the Southern hemisphere array of
CTA will consist of three types of telescopes with different
mirror sizes, 24, 10-12 and 4-6 meters, in order to cover
the full energy range. This enhanced capacity includes: a)
Higher sensitivity for source detection, b) Higher detection
rate of transient events, c) Improved angular resolution and
d) Full sky coverage. The CTA array will be distributed
over a surface of around 10 km$^2$ for the southern array, located at sites with excellent 
optical and atmospheric conditions and at an altitude
between 1800 m and 3800 m asl.

\section{Description of the proposed Argentinean sites}
 
 Argentina has many potential candidate sites satisfying the
basic requirements for installation of CTA. Mostly in the
Northwest of the country, in the provinces where the Andes mountains are placed 
(San Juan, La Rioja, Catamarca,Salta and Jujuy) it is possible to find adequate flat
areas of height above 1800 meters, with clear skies and
suitable meteorological and geographical conditions. After
a pre-selection of a dozen promising sites, the search
was successively refined by analysis of cartography, available
climate data, local surveys and interviews with experts
in the search for astronomical sites (see Allekotte et
al.\cite{bib:ale}). In particular, about the dust and sand, the requirements
are: "The natural cleanliness class of the site at 3 m
above ground should be better than ISO-Class 8 (according
to ISO14644-1) for 90\% of the time (survival condition)". Its value of 290 000 part/m$^3$ is order of magnitudes higher than we obtained for SAC (CASLEO): 610 (9750) part/m$^3$, in calm (not windy) typical days with the GRIMM instrument.

Two sites that have very good visibility (and consequently
very low aerosol content) are: San Antonio
de los Cobres (Province of Salta) and Pampa de El
Leoncito/CASLEO (Province of San Juan). They have
been preselected by the Argentinean institutional members of CTA and attention
has been put to characterize and develop them further, being
the present work a partial approach to this characterization,
concentrated in the aerosol behavior.

One of the proposed sites is very near San Antonio de
los Cobres (SAC) (24$^o$02'42.7'' S, 66$^o$14'05.8'' W) in the
high altitude desertic place of Puna of Atacama, in the
Salta Province at 3600 m a.s.l., in a rather plane region.
The other proposed site is in the Pampa (plane) of El
Leoncito (LEO) (31$^o$04'48'' S, 69$^o$16'12'' W), near the
CASLEO (Complejo Astron´omico El Leoncito), at San
Juan Province. This site is placed at 2600 m a.s.l., at the
middle of the region which is limited at the right by the
Precordillera and at the left by the Andes mountains.

\section{Results}

\subsection{Ground measurements of aerosol concentration}

\begin{figure}
\centering
\includegraphics[width=0.4\textwidth]{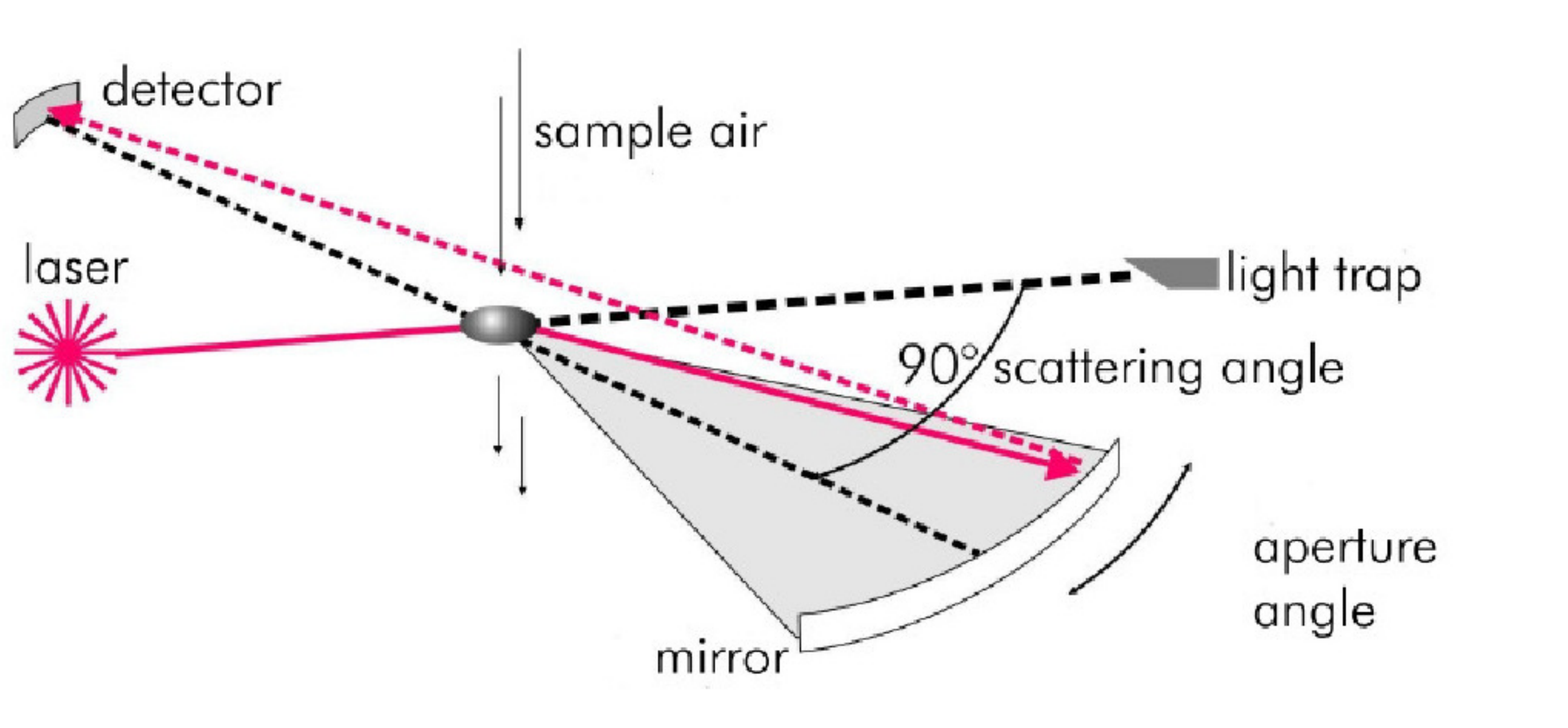}
\caption{GRIMM aerosol spectrometer, model 1.109. Schematic description of the functioning of the instrument (for more details, see the GRIMM 1109 Manual at: www.Grimm-aerosol.de).}
\label{GRIMM}
 \end{figure}

We used in the present work a GRIMM aerosol spectrometer (also called Optical particle  counter), model 1.109,
of the Institute of Physics Rosario (CONICET – National University of Rosario), Argentina, that measures the particle
concentration in  particles/m$^3$ or in $\mu$g/m$^3$. It has the possibility to measure in detail the  concentration
at  mean diameter intervals between  0.22 $\mu$m and 32 $\mu$m and at fixed (predetermined) time intervals (for
example, each 1, 5,..., minutes).  Figure \ref{GRIMM} describes its operation, that it is based in the
measurement technique of scattering of 655 nm laser light, produced by the particles that are present in the air sample. This sample is introduced into the equipment by the action of a pump with
controlled flux of 1.2 liters/min. The scattered radiation is collected by a parabolic mirror and
directed through the detector. The signal is detected by a multi-channel classifier (of 31 channels) and correlated
with the size distribution. This signal can then be transformed into aerosol number concentration or particulate matter concentration.

The GRIMM aerosol spectrometer has been used in many applications, one of the most important one was done in 2010,
during the eruption of the Eyjafjallaj\"okull Island volcano, that affected significantly the air traffic in Europe
(see for example the work of Weber et al. \cite{bib:weber7}). In what follows, we present results obtained with this
instrument in both Argentinean proposed sites. 
\\

\subsubsection{San Antonio de los Cobres data}

\begin{figure}[t]
  \centering
\includegraphics[width=0.4\textwidth]{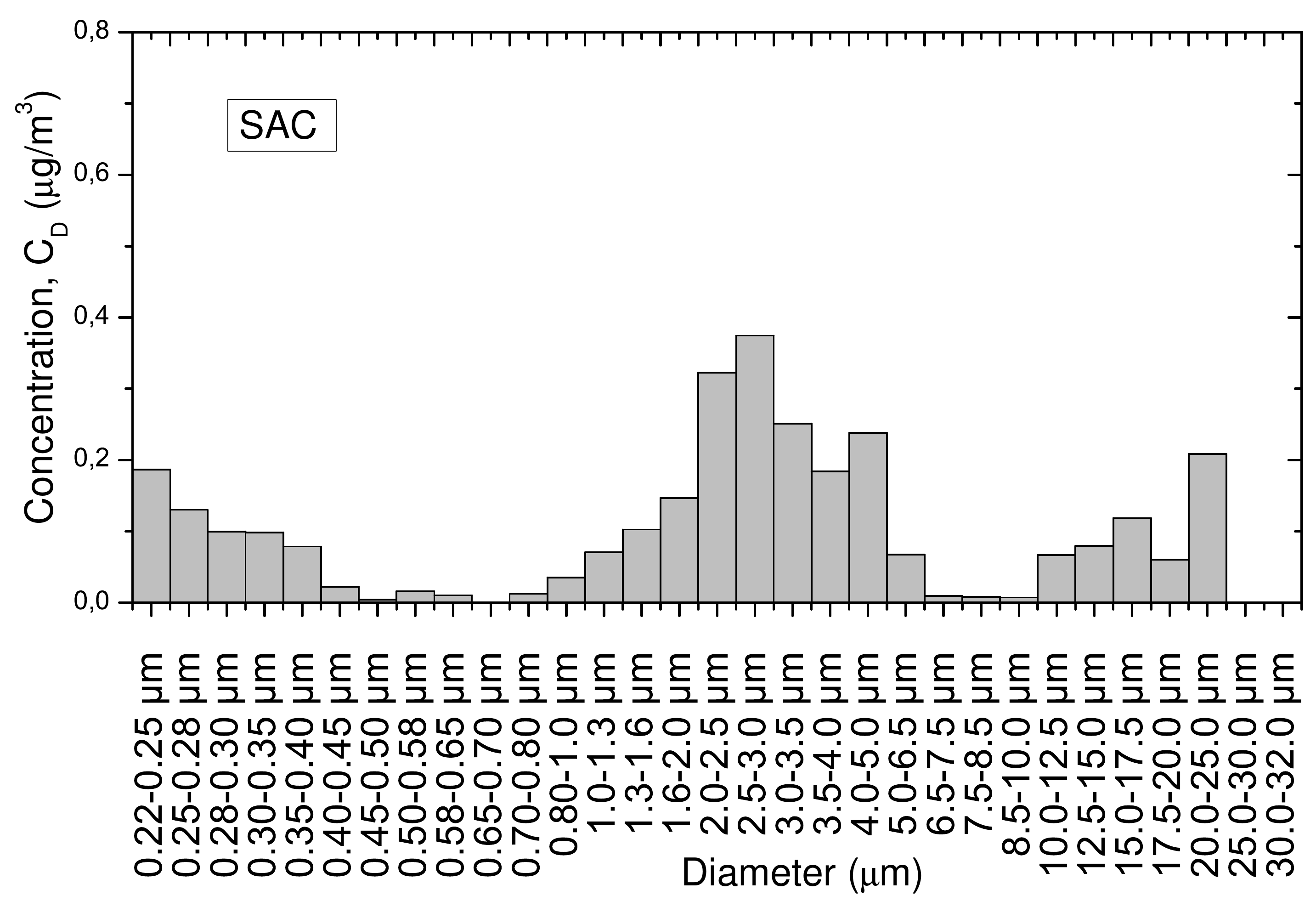} 
\caption{Aerosol concentration (in all the measured period), C$_D$, in the range 0.22-32.0 $\mu$m, for 15th December 2012 at SAC.}
\label{fig2}
\end{figure}

At the SAC location, the aerosol concentration data were collected during the morning of the day 15th December 2012,
with rather calm (not windy) conditions, being the mean wind velocity measured at this location with a Davis
meteorological station about (11 $\pm$ 7) Km/h. The aerosol concentration  shows a peak in the 2.5-3.0 $\mu$m range 
of the mean aerosol diameter and a very low mean total concentration C$_D$ = (0.097 + 0.1, -0.097) $\mu$g/m$^3$  of the aerosol diameter distribution (see Figure \ref{fig2}). 

\begin{figure}[t]
  \centering
\includegraphics[width=0.4\textwidth]{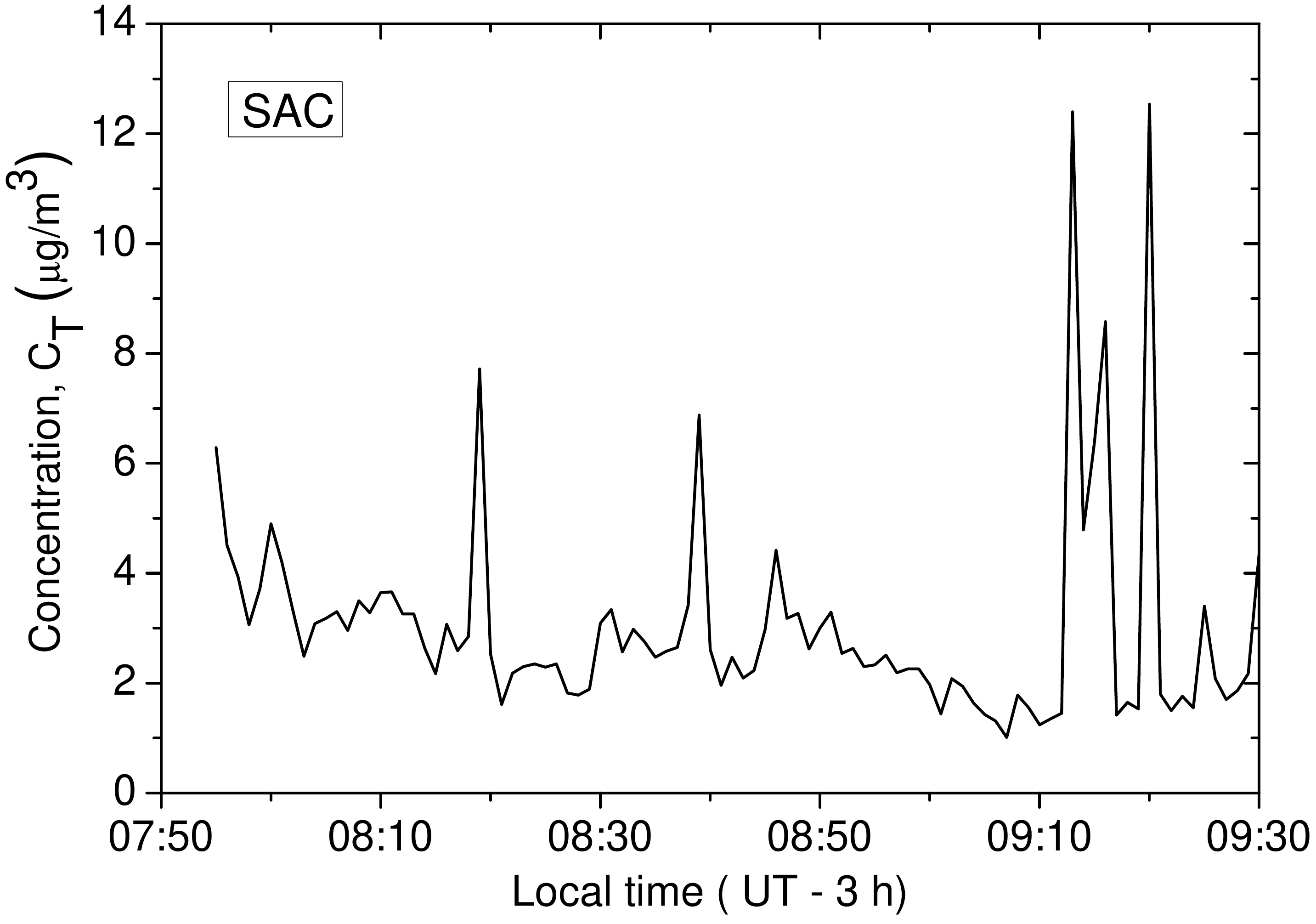} 
\caption{Aerosol concentration (for all diameters), C$_T$, in the range 0.22-32.0 $\mu$m, measured with the GRIMM  the morning of 15th December 2012, at SAC.}
\label{fig3} 
\end{figure}

We represented in Figure \ref{fig3} the same quantity but as a function of time, showing a mean value 
C$_T$ = (3.0  $\pm$ 1.9) $\mu$g/m$^3$, (where C$_T$ = C$_D$ x 31, the number of channels).  


\subsubsection{El Leoncito/CASLEO data}

\begin{figure}
  \centering
\includegraphics[width=0.4\textwidth]{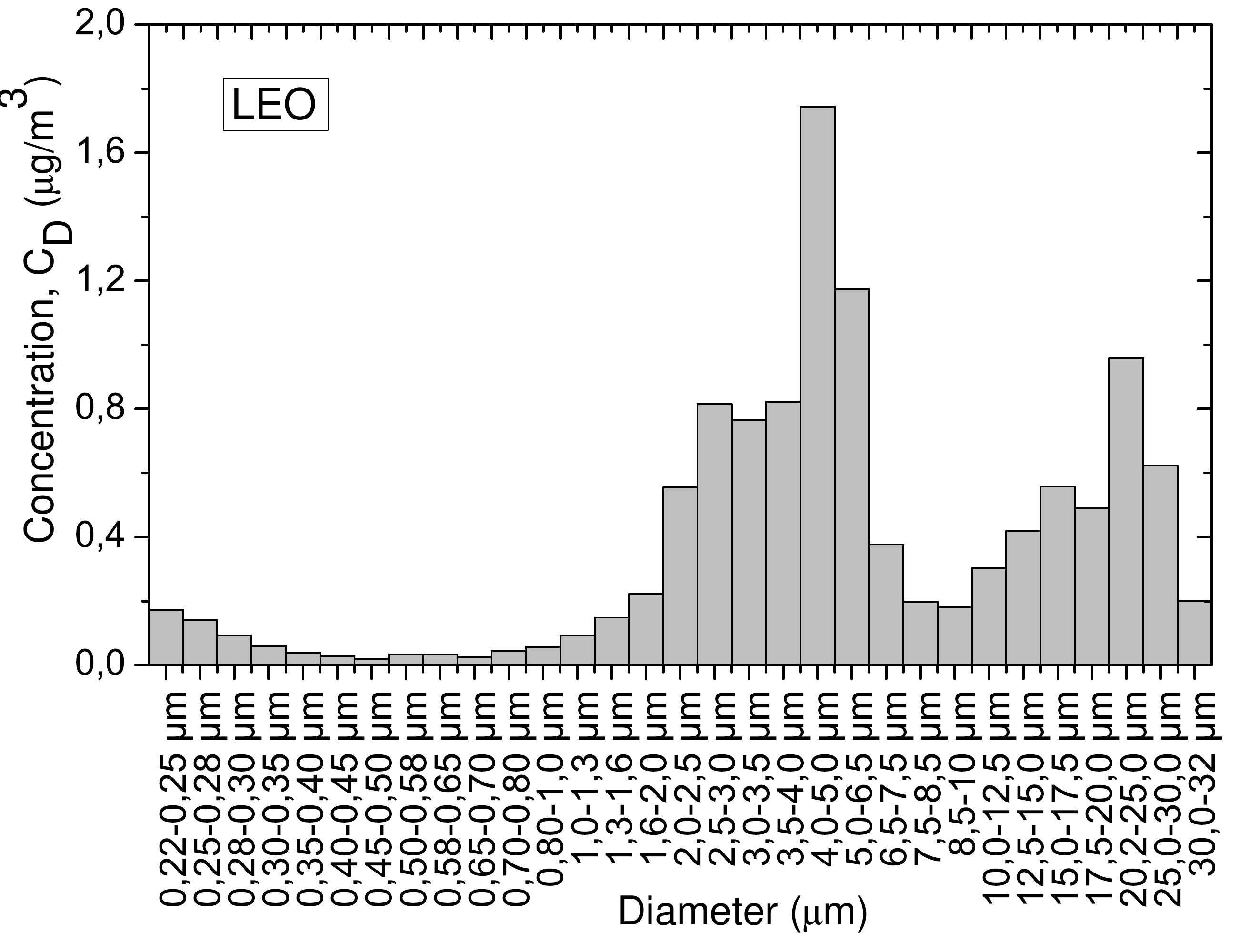} 
\caption{ Aerosol concentration  (in all the measured period), C$_D$, in the range 0.22-32.0 $\mu$m, measured with the GRIMM  from near noon  of 27th to near noon of 28th December 2012, at El Leoncito.}
\label{fig4}
\end{figure}

\begin{figure} [t]
 \centering
\includegraphics[width=0.4\textwidth]{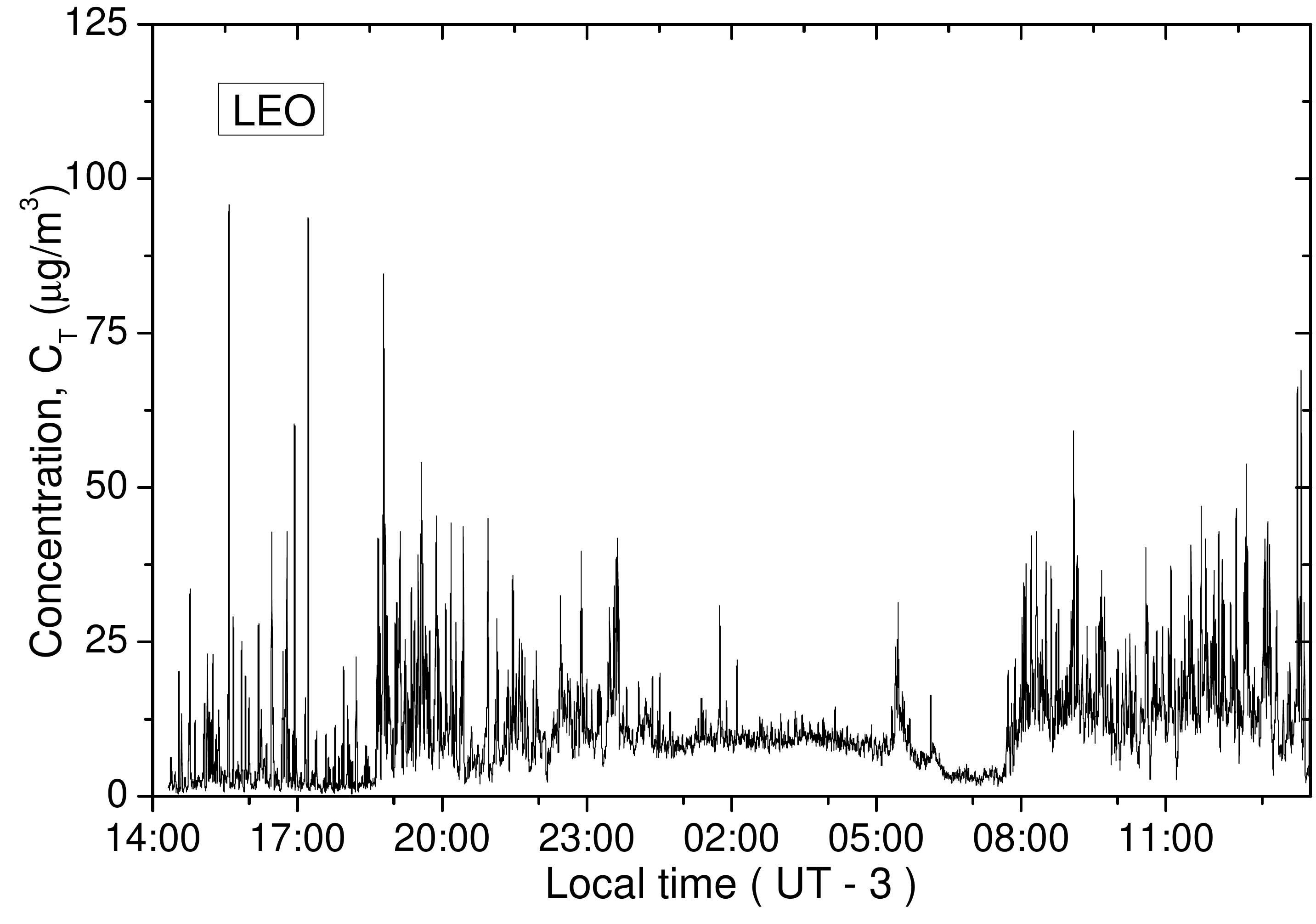} 
\caption{Aerosol concentration (for all diameters), C$_T$, in the range 0.22-32.0 $\mu$m, measured with the GRIMM, from near noon of 27th to near noon of 28th December 2012, at CASLEO}
\label{fig5}
\end{figure}

A similar analysis for the aerosol concentration was made with the GRIMM instrument at El Leoncito/CASLEO, where the
particulate matter was collected between 2 pm of the day 27th up to 2 pm of the day 28th December 2012
(Figure \ref{fig4} ). The mean wind velocity measured at this location with a Davis meteorological station
was about (10.8 $\pm$ 6) Km/h. The data show a peak in the 4.0-6.5 $\mu$m range of the mean aerosol diameter 
and a low mean total concentration  C$_D$ = (0.37 $\pm$ 0.8) $\mu$g/m$^3$. 
We represented the time dependence of this total concentration in Figure \ref{fig5}, having a mean value C$_T$ = (11.5 $\pm$ 11.4) $\mu$g/m$^3$. 
\\

\subsection{Satellite data of aerosol optical depth and 380 nm spectral irradiance}


The SeaWiFS (Sea-viewing Wide Field-of-view Sensor) instrument is on board of the SeaStar NASA spececraft, placed 
at a 705 km circular, noon, sun-synchronous orbit.  The satellite orbits the Earth fourteen times a day, every 99
minutes. It was designed mainly for the study of fitoplankton in the ocean, but it has the possibility to derive
aerosol optical depth data not only from the ocean but also from the land \cite{bib:oceancolor8}. Recently, it was
introduced the improved Deep Blue algorithm (Level 3) that gives more reliable data than the previous one \cite{bib:giovanni9}.

\subsubsection{San Antonio de los Cobres and El Leoncito/CASLEO data}

\begin{figure}
\centering 
\includegraphics[width=0.35\textwidth]{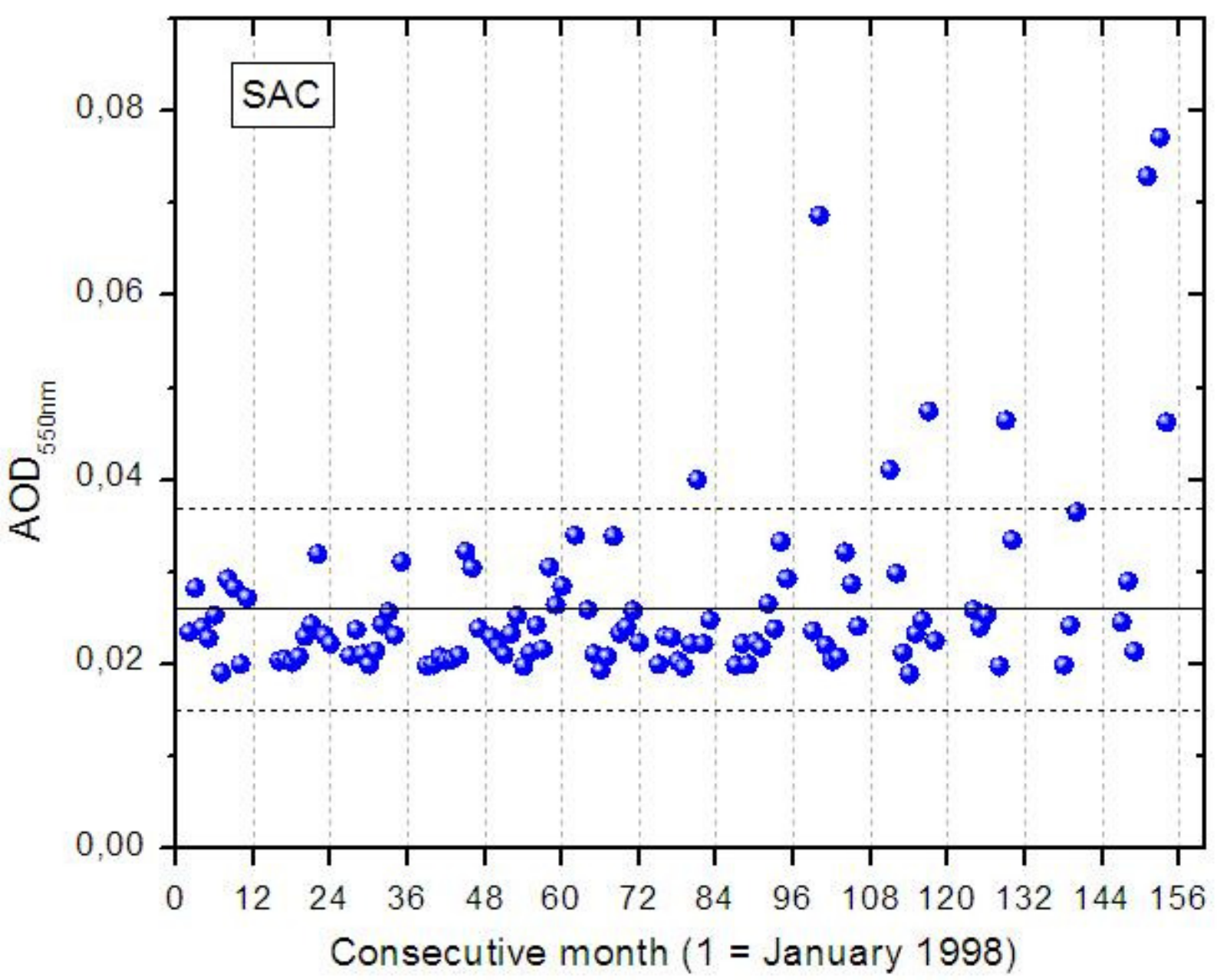} 
\caption{Area average  monthly time series (1998-2010 period) Aerosol Optical Depth at 550 nm (AOD$_{550nm}$) for SAC. The horizontal lines correspond to the mean value (solid) $\pm$ one standard deviation (dashed). Source: SeaWiFS \cite{bib:oceancolor8}.}
\label{fig6}
\end{figure}

\begin{figure}[t]
\centering 
\includegraphics[width=0.35\textwidth]{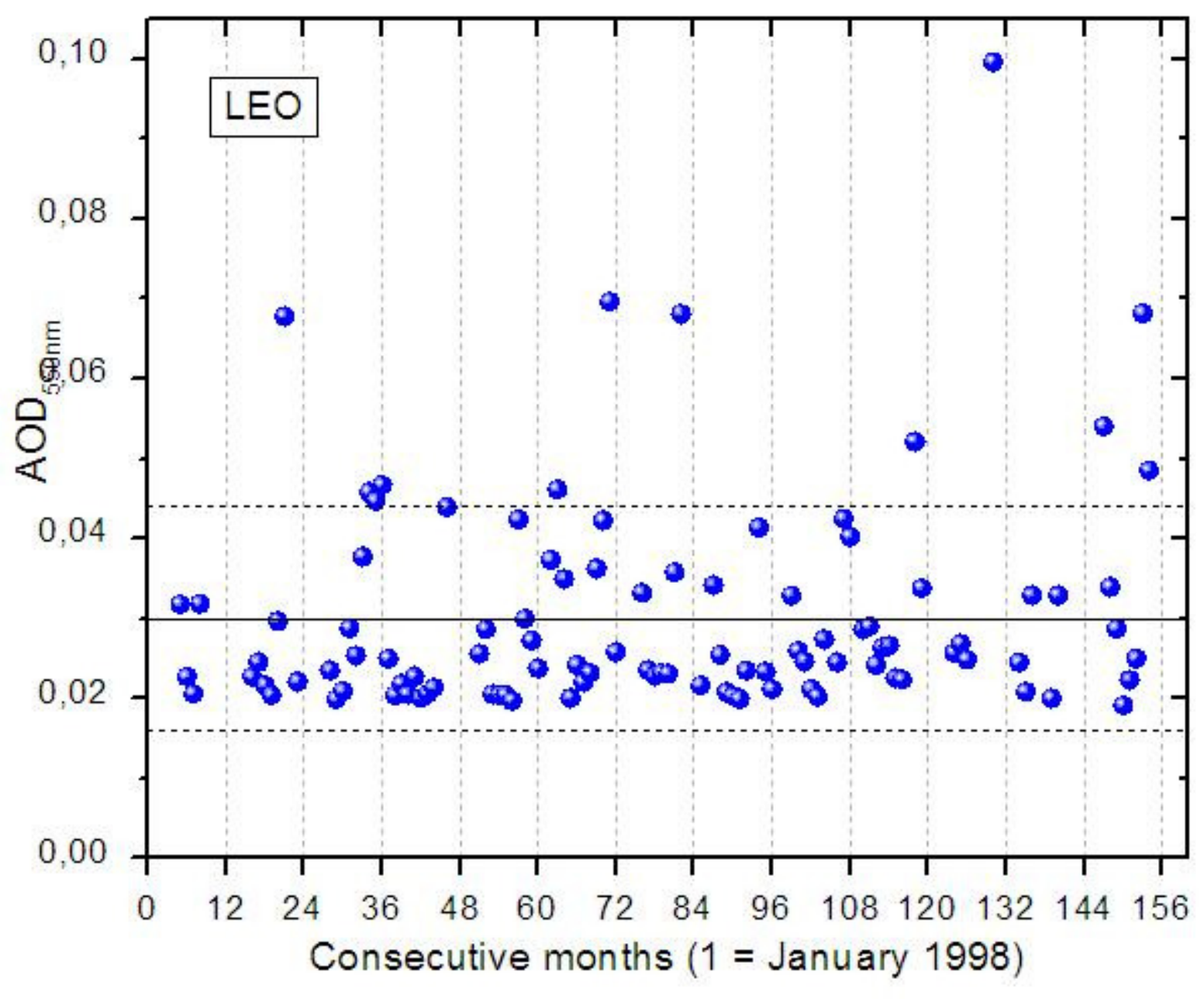} 
\caption{The same than in Figure 6 for El Leoncito/CASLEO.}
\label{fig7}
\end{figure}

The attenuation of Cherenkov photons by aerosols can be described by the aerosol optical depth  (AOD) for a given
wavelength. Consequently, we present AOD values for each CTA  Argentinean sites: SAC and  El Leoncito/CASLEO.
They are derived from data measured by the SEAWiFS instrument. The corresponding mean values for the 1998-2010 
period are very low: AOD$_{550nm,SW,SAC}$ = 0.026 $\pm$ 0.011 and AOD$_{550nm,SW,CASLEO}$ = 0.030 $\pm$ 0.014.
In Figure \ref{fig6}  and  Figure \ref{fig7} we present the mean monthly time series for the 1998-2010 period,
measured by the same instrument at SAC and El Leoncito/CASLEO sites.

\subsubsection{380 nm spectral solar irradiance}

\begin{figure}[t]
\centering
\includegraphics[width=0.5\textwidth]{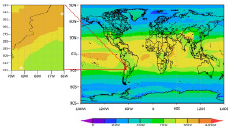} 
\caption{Spectral 380 nm - solar irradiance (in units of mW/(m$^2$nm)) measured by OMI-KNMI (The Netherlands) instrument on board of AURA/NASA satellite, in the 2005-2011 period. Note: the upper (lower) star in the figure at left indicates the position of the SAC (El Leoncito/CASLEO) site.}
\label{fig8}
\end{figure}

Atmospheric particulate matter attenuates significantly the solar irradiance, mainly in the UV (290-400 nm) and
visible (400-750 nm) ranges. Consequently, as a test in the Argentinean CTA sites of the tropospheric (0-about 
15 km height) UV atmospheric transmittance,
we introduce in Figure \ref{fig8} the mean spectral (380 nm) solar
irradiance measured by OMI-KNMI (The Netherlands) instrument
on board of AURA/NASA satellite, for the 2005-2011 
period. The corresponding value for SAC site is (925 $\pm$ 25) mW/(m$^2$nm), and for El Leoncito/CASLEO site is
(850 $\pm$ 50) mW/(m$^2$nm). It can be seen at the global false
color image (Figure 8, right), that the mean narrow band
(380 nm) solar radiation at SAC and El Leoncito-CASLEO
places is within the highest of the world. This shows a small
influence of aerosol (almost the unique large Mie scattering attenuator
in this wavelength range), and confirms the low AOD
values displayed in the previous section. \\

\section{Conclusions}

The present ground (GRIMM aerosol spectrometer) results
obtained for clear days are promising. Also from the satellite
data, we can conclude that, at least from the point of
view of the aerosol content of the atmosphere at both argentinean
proposed sites, San Antonio de los Cobres (Salta
province) and El Leoncito/CASLEO (San Juan province),
the sky is very clear and well suited for the placement of
the Cherenkov Telescope Array, since the days where typical not windy ones and the values very low, as indicated in item 3.1. In particular, about the requirements for dust and sand, its value of 290 000 part/m$^3$ is order of magnitudes higher than we obtained for SAC (CASLEO): 610 (9750) part/m$^3$, in calm (not windy) days with the GRIMM instrument.

These two sites have very good visibility (and consequently very low aerosol content).

\section{Acknowledgements}

We like to acknowledge to: the CTA Argentina-Brazil collaboration, CONICET, MYNCYT,  National University of Rosario,
National Technological University, and  SEAWiFS/NASA Science Team. Also to Christina Hsu and Andrew Sayer of
the Goddard Space Flight Center/NASA  for fruitful discussions on aerosol optical depth data. We gratefully acknowledge
support from the agencies and organizations listed in this page:\\
 http://www.cta-observatory.org/?q=node/22.

All the web pages were retrieved during may of 2013.

\end{document}